\documentclass{article}

\usepackage{arxiv}

\usepackage[utf8]{inputenc} 
\usepackage[T1]{fontenc}    
\usepackage{hyperref}       
\usepackage{url}            
\usepackage{booktabs}       
\usepackage{amsfonts}       
\usepackage{nicefrac}       
\usepackage{microtype}      
\usepackage{lipsum}		
\usepackage{graphicx}
\usepackage{multirow}
\usepackage{cite}

\usepackage{amsmath}
\usepackage{chemformula}

\usepackage{epstopdf}
\usepackage{array}
\usepackage{subfigure}
\usepackage{ragged2e}
\usepackage{booktabs,makecell,multirow,tabularx}
\usepackage[ruled,linesnumbered]{algorithm2e}

\title{NAS-PINNv2: Improved neural architecture search framework for physics-informed neural networks in low-temperature plasma simulation}

\author{Yifan Wang, ~Linlin Zhong\thanks{This paper is currently under consideration by a journal.} \\
	School of Electrical Engineering\\
	Southeast University\\
	No 2 Sipailou, Nanjing, Jiangsu Province 210096, China\\
	\texttt{linlin@seu.edu.cn}\\
}

\date{January 24, 2025}

\renewcommand{\headeright}{}
\renewcommand{\undertitle}{A Preprint}

\begin{document}
\maketitle

\begin{abstract}
Limited by the operation and measurement conditions, numerical simulation is often the only feasible approach for studying plasma behavior and mechanisms. Although artificial intelligence methods, especially physics-informed neural network (PINN), have been widely applied in plasma simulation, the design of the neural network structures still largely relies on the experience of researchers. Meanwhile, existing neural architecture search methods tailored for PINN have encountered failures when dealing with complex plasma governing equations characterized by variable coefficients and strong nonlinearity. Therefore, we propose an improved neural architecture search-guided method, namely NAS-PINNv2, to address the limitations of existing methods. By analyzing the causes of failure, the sigmoid function is applied to calculate the architecture-related weights, and a new loss term is introduced. The performance of NAS-PINNv2 is verified in several numerical experiments including the Elenbaas-Heller equation without and with radial velocity, the drift-diffusion-Poisson equation and the Boltzmann equation. The results again emphasize that larger neural networks do not necessarily perform better, and the discovered neural architecture with multiple neuron numbers in a single hidden layer imply a more flexible and sophisticated design rule for fully connected networks.
\end{abstract}

\section{INTRODUCTION}
\label{sec:sec1}
\paragraph{}
Plasma is a quasi-neutral substance composed of charged and neutral particles. Low-temperature plasma (LTP) is a type of plasma in a non-equilibrium state, where the temperature of heavy particles is rather low. LTP has numerous applications in industrial fields, such as semiconductor manufacturing \cite{cite1}, mineral processing \cite{cite2}, biomedical engineering \cite{cite3}, waste treatment \cite{cite4} and arc quenching \cite{cite5}. To investigate the mechanisms under the applications, many efforts have been made to conduct experiments or develop numerical methods on the topic of LTP \cite{cite6}. However, due to the extreme operating conditions, measurement challenges and difficulties in maintaining plasma stability, numerical simulation is often the only feasible approach to reconstructing and understanding LTP behaviors in many cases.

\paragraph{}
At the microscopic level, LTP is a collection of a large number of charged and neutral particles that collide and react with each other, while at the macroscopic level, it can be seen as a complex magnetohydrodynamic system formed by the coupling of multiple physical fields including electric fields, magnetic fields, flow fields, heat fields and force fields. In both cases, the spatiotemporal evolution of LTP can be described by a series of complex partial differential equations (PDEs). From the microscopic aspect, reaction-diffusion equations combined with Poisson equations can be used to describe the chemical reaction processes and species evolution, forming the kinetic and chemical kinetic model of LTP \cite{cite7}. From the macroscopic aspect, magnetohydrodynamic equations can be utilized to reconstruct the flow field and electromagnetic field of LTP, forming the fluid model \cite{cite8}. Regardless of the model type, the essence of numerical simulation lies on solving PDEs with given initial and boundary conditions. Typically, common numerical methods to solve PDEs in LTP models include finite element method (FEM) and finite volume method (FVM), both of which require meticulous meshing scheme when handling complex PDE solving problems, together with the efforts of equation discretization. Although many automatic meshing techniques have been developed \cite{cite9}, the limited adaptability to irregular computational domains and PDEs with strong nonlinearity still remains an obstacle to efficient numerical simulation. Especially in LTP simulation, complex dynamics such as turbulence and shock waves make it even harder to maintain a balance between computation efficiency and accuracy \cite{cite10}.

\paragraph{}
In recent years, with the development of artificial intelligence (AI) technology, trails have emerged that apply machine learning to LTP numerical simulation \cite{cite11}. Early attempts included the following pure data-driven methods such as proper orthogonal decomposition (POD), convolutional neural network (CNN), recurrent neural network (RNN) and long short-term-memory (LSTM) \cite{cite12, cite13, cite14, cite15}. Although these methods can reduce model complexity and computational cost to some extent, their essence lies in data fitting, lacking physical constraints. Physics-informed neural network (PINN), proposed by Raissi et al \cite{cite16}, provides a new approach to further address these issues. By appropriately designing a loss function, PINN can incorporate the physics information embedded in PDEs into the neural network. Following this basic idea, efforts have been made to build frameworks of LTP numerical simulation based on PINN. Focusing on Boltzmann equation, Kawaguchi et al \cite{cite17} solved the electron velocity distribution function (EVDF) based on PINN and conducted a detailed comparison with the Monte Carlo simulation (MCS) results. Further improvements have also been achieved by introducing non-dimensionalization technique \cite{cite18} and extra parameters such as reduced electric fields \cite{cite19}. Systematic study was conducted by Zhong et al \cite{cite20} on PINN-based thermal plasma simulation, investigating the application of PINN in the simulation of steady-state and transient arc discharge. Building upon this research, two LTP simulation frameworks were proposed, namely coefficient-subnet physics-informed neural network (CS-PINN) and Runge-Kutta physics-informed neural network (RK-PINN) \cite{cite21}, addressing the solution of PDEs with variable coefficient and the solution of transient PDEs with large time steps. More works based on PINN can also be found in plasma jet simulation, Vlasov equation solving and plasma turbulence simulation \cite{cite22, cite23, cite24}.

\paragraph{}
Despite the rapid development of PINN-based LTP simulation methods, the discussion on neural network architectures in this area has been overlooked. Early efforts have been made by Wang et al \cite{cite25} to explore the application of neural architecture search (NAS) in PINN and several patterns of effective neural architectures have been summarized. However, in the more complex scenario of LTP simulation with variable equation coefficients and strong nonlinearity, the design of neural architectures still requires further discussion. As a matter of fact, the previously proposed methods neural architecture search-guided PINN (NAS-PINN) does fail in LTP simulation under certain configurations. Therefore, this paper demonstrates the failure phenomenon of NAS-PINN in LTP simulation and analyzes the cause of such failure. Based on the analysis, improvements are achieved to adapt NAS-PINN to LTP simulation. Specifically, the acquisition of architecture-related weights is modified and a new loss term is introduced, constituting the improved version NAS-PINNv2. The improved method can handle the neural architecture search tasks in complex LTP simulation scenarios, exhibiting superior accuracy and robustness.

\paragraph{}
The rest of the paper is organized as follows. In Section \ref{sec:sec2}, several related works will be introduced in detail. The failure of NAS-PINN in LTP simulation will be discussed and analyzed in Section \ref{sec:sec3}, together with the improvement and the newly proposed NAS-PINNv2. Section \ref{sec:sec4} presents the numerical results of different LTP governing equations. Finally, this work will be concluded in Section \ref{sec:sec5}.

\section{RELATED WORK}
\label{sec:sec2}
\subsection{Physics-informed neural network (PINN)}
\label{sec:sec2.1}
\paragraph{}
Physics-informed neural network (PINN) was first proposed by Raissi et al. \cite{cite16} to solve forward and inverse problems of nonlinear partial differential equations (PDEs). The key idea of PINN is to impose constraints on the network output $\hat{u}$ to satisfy the equation and initial/boundary condition through a designed loss function based on the PDE. Considering a PDE in the following form:

\begin{equation}
\label{equ:equ1}
	\begin{array}{l}
  	u{{\left( t,\mathbf{x} \right)}_{t}}+\mathcal{F}\left( \mathbf{x},u\left( t,\mathbf{x} \right) \right)=f\left( t,\mathbf{x} \right),\text{       }t\in \left[ 0,T \right],\mathbf{x}\in \Omega , \\ 
 	\mathcal{B}\left( u,t,\mathbf{x} \right)=0,\text{     }x\in \partial \Omega , \\ 
 	\mathcal{I}\left( u,t,\mathbf{x} \right)=0,\text{       }t=0. \\ 
	\end{array}
\end{equation}

where $t$ and $\bf{x}$ are the temporal and spatial coordinates, respectively, $u(t, \bf{x})$ is the solution, $\mathcal{F}(\cdot)$ is the combination operator of spatial derivatives, $f(t, \bf{x})$ is the source term, $\mathcal{B}(\cdot)$ and $\mathcal{I}(\cdot)$ represent the boundary condition and initial condition, respectively.

\paragraph{}
Based on the given PDE, the loss function can be expressed as:

\begin{equation}
\label{equ:equ2}
	L={{\omega }_{\mathcal{F}}}{{L}_{\mathcal{F}}}+{{\omega }_{\mathcal{B}}}{{L}_{\mathcal{B}}}+{{\omega }_{\mathcal{I}}}{{L}_{\mathcal{I}}}
\end{equation}

\begin{equation}
\label{equ:equ3}
	{{L}_{\mathcal{F}}}=\frac{1}{{{N}_{\mathcal{F}}}}\sum\limits_{i=1}^{{{N}_{\mathcal{F}}}}{l\left( \hat{u}{{({{t}_{i}},{{\mathbf{x}}_{i}})}_{t}}+\mathcal{F}({{\mathbf{x}}_{i}},\hat{u}({{t}_{i}},{{\mathbf{x}}_{i}}))-f({{t}_{i}},{{\mathbf{x}}_{i}}) \right)}
\end{equation}

\begin{equation}
\label{equ:equ4}
	{{L}_{\mathcal{B}}}=\frac{1}{{{N}_{\mathcal{B}}}}\sum\limits_{i=1}^{{{N}_{\mathcal{B}}}}{l\left( \mathcal{B}({{{\hat{u}}}_{i}},{{t}_{i}},{{\mathbf{x}}_{i}}) \right)}
\end{equation}

\begin{equation}
\label{equ:equ5}
	{{L}_{\mathcal{I}}}=\frac{1}{{{N}_{\mathcal{I}}}}\sum\limits_{i=1}^{{{N}_{\mathcal{I}}}}{l\left( \mathcal{I}({{{\hat{u}}}_{i}},{{t}_{i}},{{\mathbf{x}}_{i}}) \right)}
\end{equation}

where $\omega_{\mathcal{F}}$, $\omega_{\mathcal{B}}$ and $\omega_{\mathcal{I}}$ are the factors to balance the corresponding parts of the loss function, $N_{\mathcal{F}}$, $N_{\mathcal{B}}$ and $N_{\mathcal{I}}$ are the numbers of collocations points in the computational domain, on the boundary and at the initial time slice, respectively, and $l(\cdot)$ is a given metric function such as $L^1$ or $L^2$ norm.

\paragraph{}
By minimizing the designed loss function, the output of PINN $\hat{u}$ can finally converge to the true solution of the given PDE. Since the spatial and temporal derivatives can be obtained through automatic differentiation \cite{cite26} (AD) in deep learning frameworks, PINN can be conveniently implemented in various circumstances. It has been successfully tailored and employed to solve Navier-Stokes equation \cite{cite27}, Schrodinger equation \cite{cite16}, Helmholtz equation \cite{cite28} and other problems involving PDEs.

\subsection{Coefficient-Subnet PINN (CS-PINN) and Runge-Kutta PINN (RK-PINN)}
\label{sec:sec2.2}
\paragraph{}
In the realm of LTP simulation, Zhong et al. proposed coefficient-subnet PINN (CS-PINN) and Runge-Kutta PINN (RK-PINN) \cite{cite21} to solve plasma governing equations. The most significant distinction of plasma governing equations is that the equation coefficients, denoted as $\bf{\lambda}$, are dependent on the equation solutions themselves, which brings much complexity to the solving process.

\begin{figure}
	\centering
	\includegraphics[width=15cm]{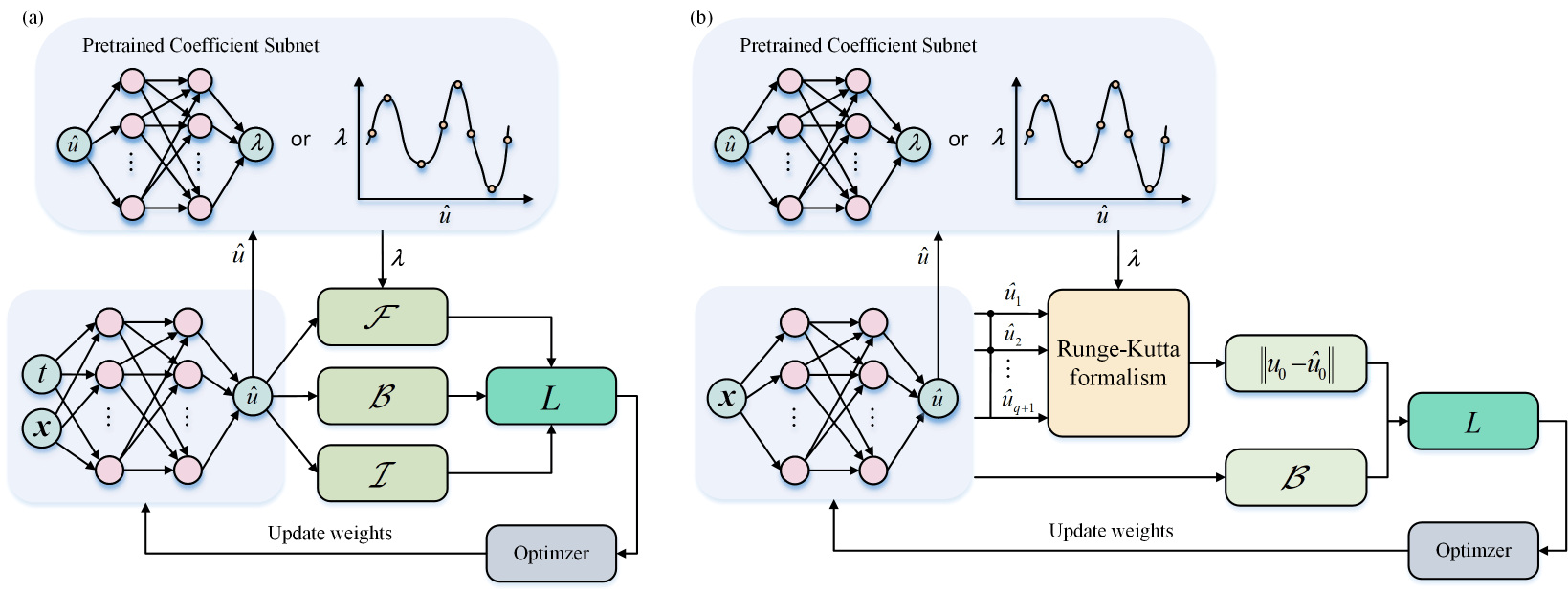}
	\caption{Framework of (a) CS-PINN and (b) RK-PINN}
	\label{fig:fig1}
\end{figure}

\paragraph{}
To tackle this problem, CS-PINN introduces an auxiliary subnet, called coefficient-subnet, as a surrogate model of the relationship between the equation coefficients and the equation solutions, as shown in Figure \ref{fig:fig1} (a). The coefficient-subnet can be either a pretrained small-scale neural network or an interpolation function, depending on the degree of nonlinearity of the coefficients and the order of differentiation in the equations. Generally, for a plasma governing equation with no more than two-order differential operators, a coefficient-subnet represented by a cubic spline function is recommended.

\paragraph{}
RK-PINN is tailored for solving plasma transient equations and allows a large time step $\Delta t$ by introducing the implicit Runge-Kutta time-stepping scheme with $q$ stages, which can be expressed as: \cite{cite29}

\begin{equation}
\label{equ:equ6}
	\begin{array}{l}
		u\left( {{t}_{0}}+\Delta t\cdot {{c}_{i}} \right)=u\left( {{t}_{0}} \right)+\Delta t\sum\limits_{j=1}^{q}{{{a}_{ij}}\left( f\left( {{t}_{0}}+\Delta t\cdot {{c}_{j}} \right)-\mathcal{F}\left( u\left( {{t}_{0}}+\Delta t\cdot {{c}_{j}} \right) \right) \right)},\text{    }i=1,2,...,q \\ 
 		u\left( {{t}_{0}}+\Delta t \right)=u\left( {{t}_{0}} \right)+\Delta t\sum\limits_{j=1}^{q}{{{b}_{j}}\left( f\left( {{t}_{0}}+\Delta t\cdot {{c}_{j}} \right)-\mathcal{F}\left( u\left( {{t}_{0}}+\Delta t\cdot {{c}_{j}} \right) \right) \right)} \\
	\end{array}
\end{equation}

where $a_{ij}$, $b_{j}$ and $c_{i}$ are the parameters of the implicit Runge-Kutta scheme.

\paragraph{}
Instead of directly outputting the solutions to the equations, RK-PINN outputs $q+1$ elements corresponding to the solutions at $q+1$ timesteps in the Runge-Kutta scheme, as displayed in Figure \ref{fig:fig1} (b). The loss function of RK-PINN can be similarly designed as Eq. \ref{equ:equ2, equ:equ3, equ:equ4, equ:equ5}, except for the term $L_{\mathcal{F}}$, which should be rewritten as:

\begin{equation}
\label{equ:equ7}
	{{L}_{\mathcal{F}}}=\frac{1}{{{N}_{\mathcal{F}}}}\sum\limits_{i=1}^{{{N}_{\mathcal{F}}}}{\sum\limits_{j=1}^{q+1}{\left\| u_{0}^{i}-\hat{u}_{0}^{i,j} \right\|}}
\end{equation}

where $u_{0}^{i}$ is the given initial value and $\hat{u}_{0}^{j}$ is the predicted initial value, which can be calculated based on Eq. \ref{equ:equ6}.

\paragraph{}
The effectiveness of CS-PINN and RK-PINN has been demonstrated in Zhong’s work \cite{cite21} in detail. Although previous studies briefly analyzed the impact of the number of layers and neurons, the design of neural architectures still lacks more specific guidance, which will be revealed in following sections.

\subsection{Neural Architecture Search-guided PINN (NAS-PINN)}
\label{sec:sec2.3}
\paragraph{}
NAS-PINN was proposed by Wang et al \cite{cite25} to search for the best neural architecture for solving a specific PDE. Due to the discrete nature of the neural network architecture design process, a relaxation is introduced to make the discrete search space continuous, which can be expressed as: \cite{cite30}

\begin{equation}
\label{equ:equ8}
	{{\bar{o}}^{(i,j)}}(x)=\sum\limits_{o\in O}{\frac{\exp (\alpha _{o}^{(i,j)})}{\sum\limits_{o'\in O}{\exp (\alpha _{o'}^{(i,j)})}}o(x)}
\end{equation}

where $\bar{o}^{(i,j)}(x)$ is the relaxed operation between the $i$-th layer and the $j$-th layer, $\alpha_{o}^{(i,j)}$ is the weight of operation $o$ in candidate operation set $O$. Applying Eq. \ref{equ:equ8}, the best neural network operations can be selected based on the optimized operation weights $\alpha_{o}^{(i,j)}$ with a certain gradient-based optimization algorithm.

\paragraph{}
Furthermore, to deal with the problem of searching for number of neurons, which is impractical in Eq. \ref{equ:equ8} as a result of the rules of tensor addition, one-zero tensor masks are incorporated in NAS-PINN. Masks are tensors with a shape of 1×$k$, where the first $j$ elements are 1 and the remaining elements are 0. By applying different values to $j$, masks can make approximations to tensors in different shapes. Consequently, the relaxation operation in NAS-PINN can be obtained as:

\begin{equation}
\label{equ:equ9}
	\bar{o}(x)={{\alpha }_{1}}\cdot \mathbf{x}+{{\alpha }_{2}}\cdot \sigma (\mathbf{w}\cdot \mathbf{x}+\mathbf{b})\cdot {{\left( \left[ {{g}_{1}},{{g}_{2}},...,{{g}_{n}} \right]\times \left[ \begin{array}{l}
   \mathbf{mas}{{\mathbf{k}}_{1}} \\ 
  \mathbf{mas}{{\mathbf{k}}_{2}} \\ 
  \text{   }... \\ 
  \mathbf{mas}{{\mathbf{k}}_{n}} \\ 
\end{array} \right] \right)}^{T}}
\end{equation}

where $\alpha_1$ is the weight for identity transformation which represents omitting this layer (referred to as skip operation), $\alpha_2$ is the weight for reserving this layer, $g$ is the weight for the number of neurons, $\mathbf{mask}$ is the one-zero tensor mask, $\mathbf{w}$ and $\mathbf{b}$ are the weights and bias of one linear hidden layer. It is worth noting that $\alpha$ and $g$ are softmax-weighted, respectively.

\paragraph{}
Based on Eq. \ref{equ:equ9}, a relaxed dense neural network (DNN) can be achieved. As shown in Eq. \ref{equ:equ10}, the relaxed DNN can be optimized through a bi-level optimization where the inner loop optimizes the weights and bias of the linear layers based on the PINN loss function, and the outer loop optimizes the weights $\alpha$ and $g$ based on a mean-square error (MSE) loss between predicted $\hat{u}$ and label values $u$. When the optimization converges, the number of hidden layers will be decided first based on $\alpha$ and then the number of neurons for each layer based on $g$.

\begin{equation}
\label{equ:equ10}
	\begin{array}{l}
   \underset{\alpha ,g}{\mathop{\min }}\,MSE\left( \hat{u},u,{{\mathbf{w}}^{*}},{{\mathbf{b}}^{*}},\alpha ,g \right) \\ 
  s.t.\text{   }({{\mathbf{w}}^{*}},{{\mathbf{b}}^{*}})=\underset{\mathbf{w},\mathbf{b}}{\mathop{\arg \min }}\,L\left( \mathbf{w},\mathbf{b},\alpha ,g \right) \\ 
\end{array}
\end{equation}

\section{METHOD}
\label{sec:sec3}
\subsection{Failure of NAS-PINN}
\label{sec:sec3.1}
\paragraph{}
Although NAS-PINN has been verified on a bundle of PDEs such as Poisson equation, Advection equation and Burgers equation \cite{cite25}, a significant performance degradation is observed when it is utilized on complex LTP governing equations.

\paragraph{}
In previous works, NAS-PINN has already demonstrated a preference for residual structures when solving PDEs with strong nonlinearity or in irregular computational domains \cite{cite25}. It was argued that reserving residual connections can resemble residual modules as in ResNet \cite{cite31}, which will raise the neural network performance on both convergence and accuracy. However, the performance raise becomes a hindrance to exploring the optimal neural architecture in the search space when the problem scenario becomes more complex. As illustrated in Figure \ref{fig:fig5}, when solving an Elenbaas-Heller equation for 1D arc plasma, NAS-PINN assigns extremely large weights to skip operations. Consequently, NAS-PINN fails to identify any feasible neural architectures, as all layers are omitted, and this outcome is evidently unreasonable.

\paragraph{}
Similar phenomenon have also been reported by Chu’s work \cite{cite32} and it is summarized as “exclusive competition”. Since the architecture-related weights in NAS-PINN are obtained using a softmax function, it is inherently constrained that the sum of all the weights equals 1. Therefore, the performance improvement introduced by residual modules results in a preference for skip operations, thereby suppressing other operations and causing the failure of NAS-PINN in complex problem scenarios.

\subsection{NAS-PINNv2}
\label{sec:sec3.2}
\paragraph{}
To mitigate such exclusive competition of skip operations, the acquisition of architecture-related weights is modified and a new loss term is introduced into the framework of NAS-PINN, constituting NAS-PINNv2.

\paragraph{}
In NAS-PINNv2, skip operation is considered at the same level as the number of neurons, which means $\alpha_2$ in Eq. \ref{equ:equ9} is removed and the relaxation operation in NAS-PINNv2 becomes:

\begin{equation}
\label{equ:equ11}
	\bar{o}(x)={{g}_{0}}\cdot \mathbf{x}+\sigma (\mathbf{w}\cdot \mathbf{x}+\mathbf{b})\cdot {{\left( \left[ {{g}_{1}},{{g}_{2}},...,{{g}_{n}} \right]\times \left[ \begin{array}{l}
   \mathbf{mas}{{\mathbf{k}}_{1}} \\ 
  \mathbf{mas}{{\mathbf{k}}_{2}} \\ 
  \text{   }... \\ 
  \mathbf{mas}{{\mathbf{k}}_{n}} \\ 
\end{array} \right] \right)}^{T}}
\end{equation}

where $g_0$ stands for the weight of skip operation and $g_1$ to $g_n$ stand for the weights of different neuron numbers. All the values of $g$ are weighted by a sigmoid function instead of a softmax function, as expressed in Eq. \ref{equ:equ12}. By applying the sigmoid function, the range of the architecture-related weights is maintained between 0 and 1, but the suppression among them is eliminated since the sum is no longer constrained to equal 1.

\begin{equation}
\label{equ:equ12}
	{{g}_{i}}=sigmoid({{G}_{i}})=\frac{1}{1+\exp (-{{G}_{i}})},\text{    }i=1,2,...,n
\end{equation}

\paragraph{}
Moreover, it is expected that the architecture-related weights will be optimized to converge to either 0 or 1, so that a determined neural architecture can be derived. Based on this requirement, a new differentiable loss term expressed as Eq. \ref{equ:equ13}, which has a global maximum at 0.5 and a global minimum at 0 and 1, is introduced \cite{cite32}. The global maximum at 0.5 makes a reasonable initial value for g since the sigmoid function will be 0.5 when it takes an input of 0. The global minimum at 0 and 1 drives the algorithm to determine an exact decision for each operation when converged, which can not only help the derivation of final neural architecture, but also make the search space more similar to the discrete neural architecture derived.

\begin{equation}
\label{equ:equ13}
	{{L}_{0-1}}=-\frac{1}{n}\sum\limits_{i=0}^{n}{{{({{g}_{i}}-0.5)}^{2}}}
\end{equation}

\begin{figure}
	\centering
	\includegraphics[width=12cm]{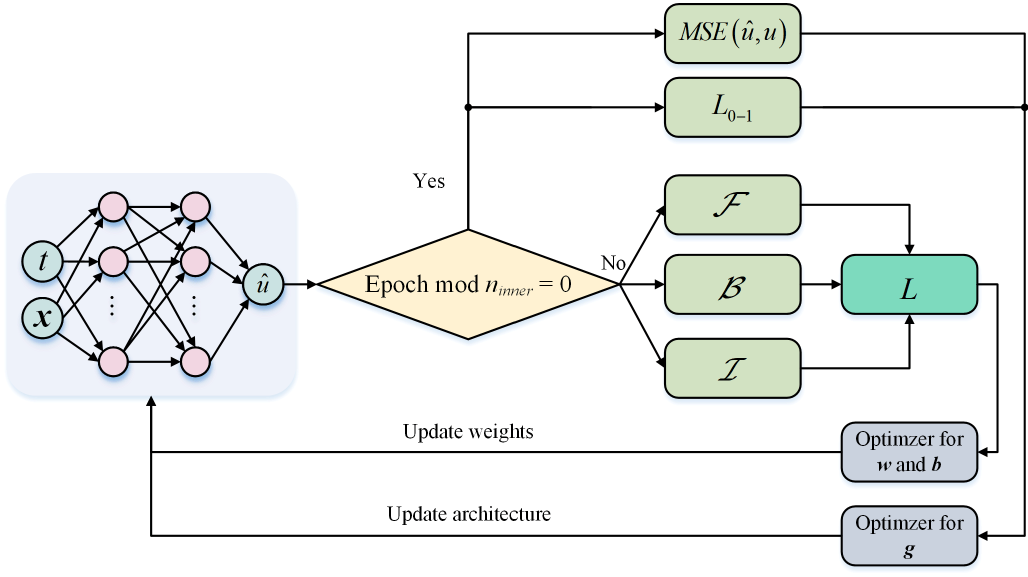}
	\caption{Framework of NAS-PINNv2}
	\label{fig:fig2}
\end{figure}

\paragraph{}
Thereby, the optimization goal of NAS-PINNv2 can be rewritten as:

\begin{equation}
\label{equ:equ14}
	\begin{array}{l}
   \underset{g}{\mathop{\min }}\,\left( MSE\left( \hat{u},u,{{\mathbf{w}}^{*}},{{\mathbf{b}}^{*}},g \right)+{{\omega }_{0-1}}{{L}_{0-1}}\left( {{\mathbf{w}}^{*}},{{\mathbf{b}}^{*}},g \right) \right) \\ 
  s.t.\text{   }({{\mathbf{w}}^{*}},{{\mathbf{b}}^{*}})=\underset{\mathbf{w},\mathbf{b}}{\mathop{\arg \min }}\,L\left( \mathbf{w},\mathbf{b},g \right) \\ 
	\end{array}
\end{equation}

where $\omega_{0-1}$ is the weight for loss term $L_{0-1}$. The process is still a bi-level optimization problem solved through an alternate optimization, demonstrated in Figure \ref{fig:fig2} and Algorithm \ref{code:alg1}.

\section{NUMERICAL EXPERIMENTS}
\label{sec:sec4}
\paragraph{}
In this section, several numerical results of plasma PDEs are discussed, including the Elenbaas-Heller equation for 1D arc plasma, the drift-diffusion-Poisson equation for 1D DC corona discharge plasma and the Boltzmann equation for electron transport. The neural architectures searched by NAS-PINNv2 are analyzed and comparisons between NAS-PINNv2 and vanilla NAS-PINN are presented.

\begin{algorithm}[htbp]
	\caption{NAS-PINNv2}
		create a DNN whose hidden layers are based on Eq. (\ref{equ:equ11})\\
		\label{code:alg1}
		set the number of epochs $n_{outer}$ for the outer loop and the number of inner loops $n_{inner}$ in one outer loop\\
		\textbf{while} $epoch < n_{outer}$ \textbf{do}\\
		~~update $\mathbf{w}$ and $\mathbf{b}$ using $L$ expressed in Eq. (\ref{equ:equ2})\\
		~~\textbf{if} $epoch$ mod $n_{inner} == 0$\\
		~~~~update $g$ using loss function expressed in Eq. (\ref{equ:equ14})\\
		~~$epoch = epoch+1$\\
		derive the discrete neural architecture according to $g$\\
\end{algorithm}

\subsection{Elenbaas-Heller equation for 1D arc plasma without radial velocity}
\label{sec:sec4.1}
\paragraph{}
Arc plasma is a type of thermal plasma generated by an electric arc, which forms when a high current flows through a gas medium. It is widely applied in industries such as welding, energy generation and plasma spraying. Following Zhong’s work \cite{cite33}, the Elenbaas-Heller equation is introduced to describe the decaying characteristics of arc plasma. Considering a cylindrically symmetric arc plasma under local thermodynamic equilibrium (LTE) and neglecting radial velocity, the 1D arc model can be expressed as:

\begin{equation}
\label{equ:equ15}
	\rho {{C}_{p}}\frac{\partial T}{\partial t}=\frac{1}{r}\frac{\partial }{\partial r}\left( r\kappa \frac{\partial T}{\partial r} \right)+\sigma \frac{{{I}^{2}}}{{{g}_{c}}^{2}}-{{E}_{rad}}
\end{equation}

where $\rho$ is the mass density, $C_p$ is the specific heat, $T$ is the arc temperature, $r$ is the radial distance, $\kappa$ is the thermal conductivity, $\sigma$ is the electrical conductivity, $I$ is the current which is set to 200 A at $t=0$ and then 0 after that, $g_c$ is the arc conductance and $E_{rad}$ is the energy loss due to the radiation transport which is described through a net emission model \cite{cite34}. The arc conductance $g_c$ can be calculated by:

\begin{equation}
\label{equ:equ16}
	{{g}_{c}}=\int_{0}^{R}{2\pi r\sigma dr}
\end{equation}

where $R$ is the radius of the arc.

\paragraph{}
In the Elenbaas-Heller equation, $\rho$, $C_p$, $\kappa$, $\sigma$ and $E_{rad}$ are all dependent on the arc temperature, thus CS-PINN is applied here to model the relationship between temperature and these properties \cite{cite21}. The initial condition is obtained by solving a steady-state Elenbaas-Heller equation in the same form as Eq. \ref{equ:equ15} without the temporal derivative term. As for the boundary condition, the temperature gradient is set to 0 at the symmetric center ($r=0$) and the temperature is set to a constant value of 2000 K at the cylindrical wall ($r=R$).

\begin{figure}
	\centering
	\includegraphics[width=17cm]{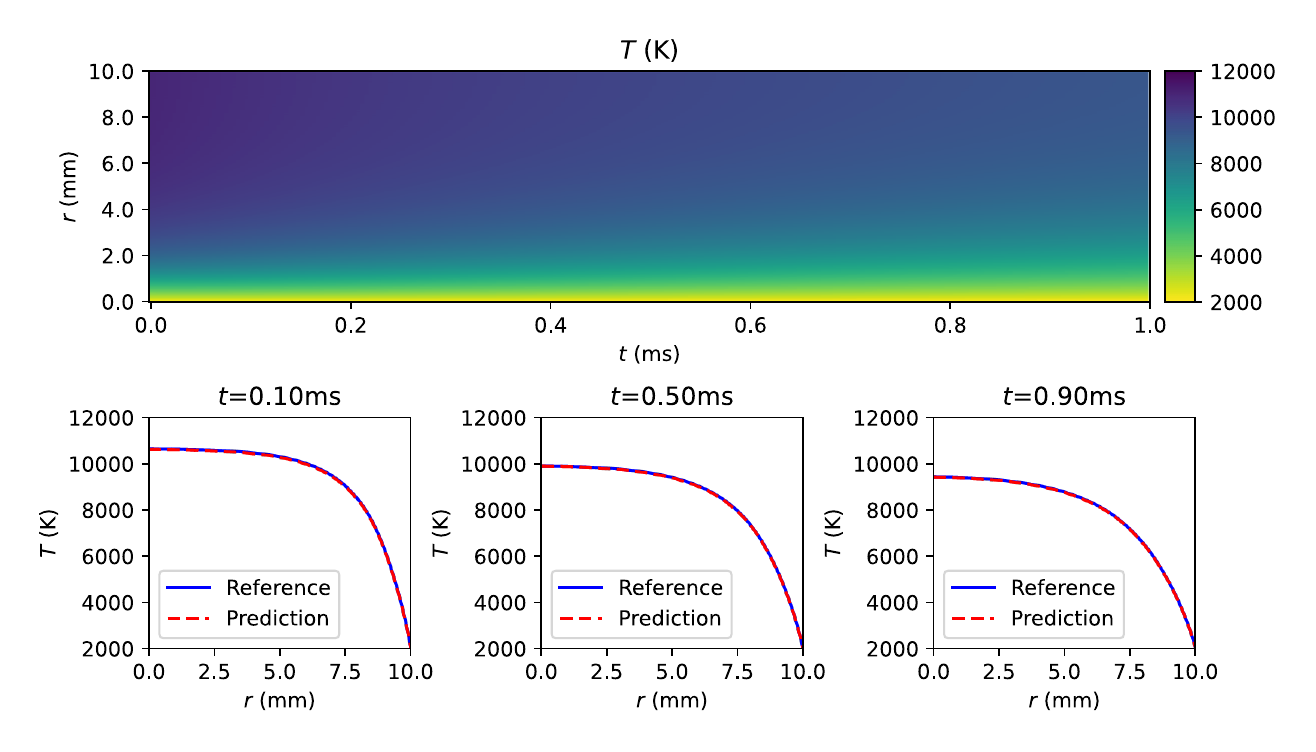}
	\caption{Prediction of the arc temperature in arc plasma of Ar by NAS-PINNv2}
	\label{fig:fig3}
\end{figure}

\begin{figure}
	\centering
	\includegraphics[width=17cm]{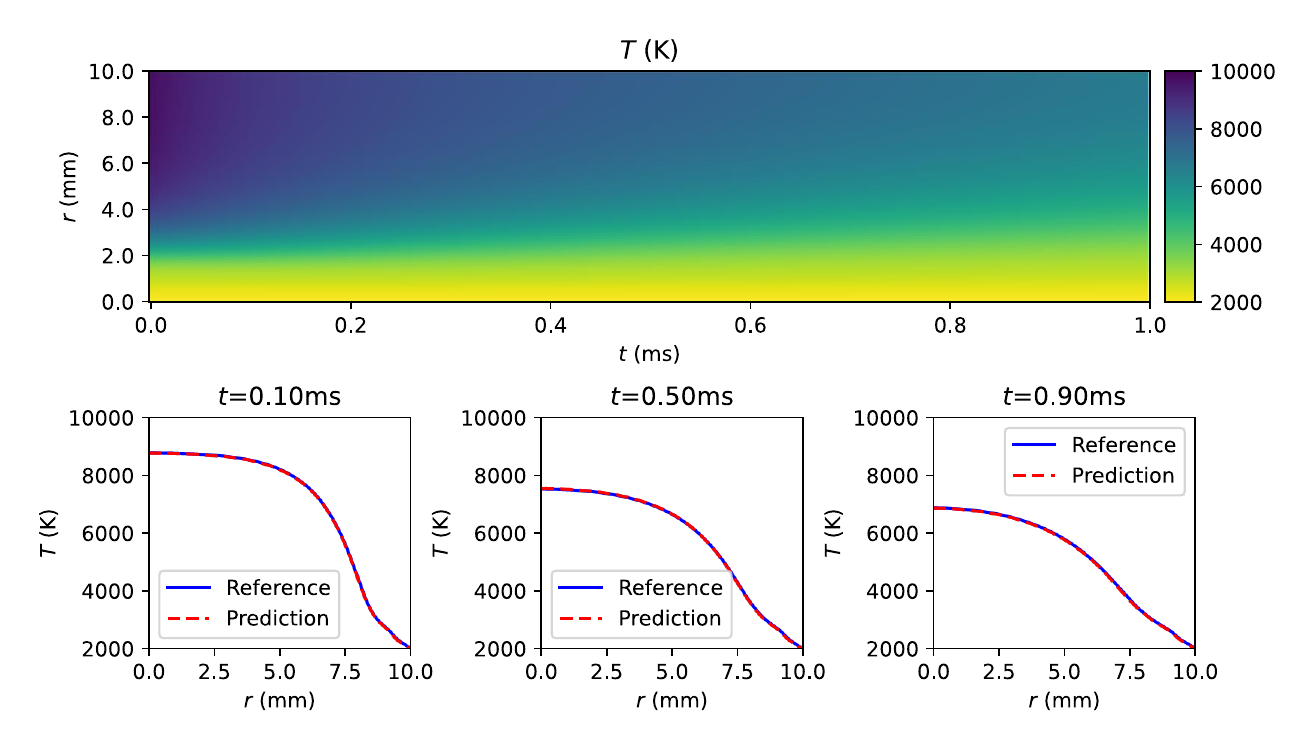}
	\caption{Prediction of the arc temperature in arc plasma of SF$_6$ by NAS-PINNv2}
	\label{fig:fig4}
\end{figure}

\paragraph{}
In this case, the framework of CS-PINN is applied to solve the arc plasmas of Ar, SF$_6$ and N$_2$. The radius of the arc $R$ is set to 10 mm, and the total evolution time duration is 1 ms. The search space is a neural network with a maximum number of 8 hidden layers, and the numbers of neurons each layer can be selected from 20 to 200 in increments of 20. Both the proposed NAS-PINNv2 and the vanilla NAS-PINN are investigated in the same search space. 200 uniformly distributed points along the radial axis and 100 uniformly distributed points along the timeline are used for inner loop training. For the architecture search phase, 1000 points at 3 selected time steps are sampled. To compare with the discovered neural architecture by NAS-PINNv2 and NAS-PINN, the largest neural network in the search space is separately selected and trained for reference. All the neural architectures are trained from scratch before tested and the reference solutions come from a high-order boundary value problem (BVP) solver \cite{cite35}. Figure \ref{fig:fig3} and Figure \ref{fig:fig4} illustrate the predicted solutions in Ar and SF$_6$, respectively, and the relative $L^2$ error are listed in Table \ref{tab:tab1}.

\begin{table}[htbp]
	\centering
	\caption{Elenbaas-Heller equation without radial velocity: relative $L^2$ error in different gases}
	\label{tab:tab1}
	\begin{tabular}{|l|l|l|l|l|}
\hline
Gas                  & Architecture name & Architecture                                         & Relative $L^2$ error &  \\ \hline
\multirow{3}{*}{Ar}  & NAS-PINNv2        & {[}2, 160, 160, 80, 140, 40, 120, 20, 1{]}           & $4.72\times10^{-5}$         &  \\ \cline{2-5} 
                     & NAS-PINN          & {[}2, 160, 140, 160, 160, 180, 200, 140,   180, 1{]} & $2.10\times10^{-5}$         &  \\ \cline{2-5} 
                     & Giant             & {[}2, 200$\times$8, 1{]}                                    & $4.58\times10^{-4}$         &  \\ \hline
\multirow{3}{*}{SF$_6$} & NAS- PINNv2       & {[}2, 120, 20, 40, 20, 120, 40, 180, 1{]}            & $3.58\times10^{-4}$         &  \\ \cline{2-5} 
                     & NAS-PINN          & {[}2, 120, 140, 60, 1{]}                             & $6.44\times10^{-4}$         &  \\ \cline{2-5} 
                     & Giant             & {[}2, 200$\times$8, 1{]}                                    & $3.86\times10^{-4}$         &  \\ \hline
\multirow{3}{*}{N$_2$}  & NAS-PINNv2        & {[}2, 40, 20, 20, 40, 120, 60, 120, 1{]}             & $7.77\times10^{-5}$         &  \\ \cline{2-5} 
                     & NAS-PINN          & {[}2, 160, 140, 160, 160, 200, 140, 180, 1{]}        & $4.39\times10^{-4}$         &  \\ \cline{2-5} 
                     & Giant             & {[}2, 200$\times$8, 1{]}                                    & $4.46\times10^{-4}$         &  \\ \hline
	\end{tabular}
\end{table}

\paragraph{}
The results demonstrate that both NAS-PINNv2 and NAS-PINN can discover a neural architecture with better performance than Giant. Since the radial velocity term is neglected in this case, reducing the complexity of the Elenbaas-Heller equation, NAS-PINN has not yet encountered failure here. However, NAS-PINNv2 can achieve a smaller relative $L^2$ error compared to NAS-PINN, except for the case in Ar. This may be a result of the elimination of suppression among the architecture-related weights computed via sigmoid function, leading to a certain discrepancy between the relaxed search space and the discrete neural architecture. Nevertheless, the relative $L^2$ error of NAS-PINNv2 is still an order of magnitude lower than Giant in the case of Ar. Generally, NAS-PINNv2 can outperform NAS-PINN in a relatively simple scenario. However, as the problem becomes more complex, which will be investigated in the next case, NAS-PINNv2 can further demonstrate its advantages.

\subsection{Elenbaas-Heller equation for 1D arc plasma with radial velocity}
\label{sec:sec4.2}
\paragraph{}
In this case, an extra radial velocity term $v_r$ is introduced, and the Elenbaas-Heller equation can be rewritten as:

\begin{equation}
\label{equ:equ17}
	\begin{array}{l}
   \frac{\partial \rho }{\partial t}+\frac{1}{r}\frac{\partial }{\partial r}\left( r\rho {{v}_{r}} \right)=0, \\ 
  \rho {{C}_{p}}\left( \frac{\partial T}{\partial t}+{{v}_{r}}\frac{\partial T}{\partial r} \right)=\frac{1}{r}\frac{\partial }{\partial r}\left( r\kappa \frac{\partial T}{\partial r} \right)+\sigma \frac{{{I}^{2}}}{{{g}_{c}}^{2}}-{{E}_{rad}} \\ 
\end{array}
\end{equation}

\paragraph{}
Both the CS-PINN framework and the RK-PINN framework are considered in this case. The search space is a neural network with a maximum number of 5 hidden layers, and the candidate neuron numbers each layer can be selected from 50 to 500 in increments of 50. The settings of the arc model remain the same as in Sec. \ref{sec:sec4.1}.

\begin{figure}
	\centering
	\includegraphics[width=17.2cm]{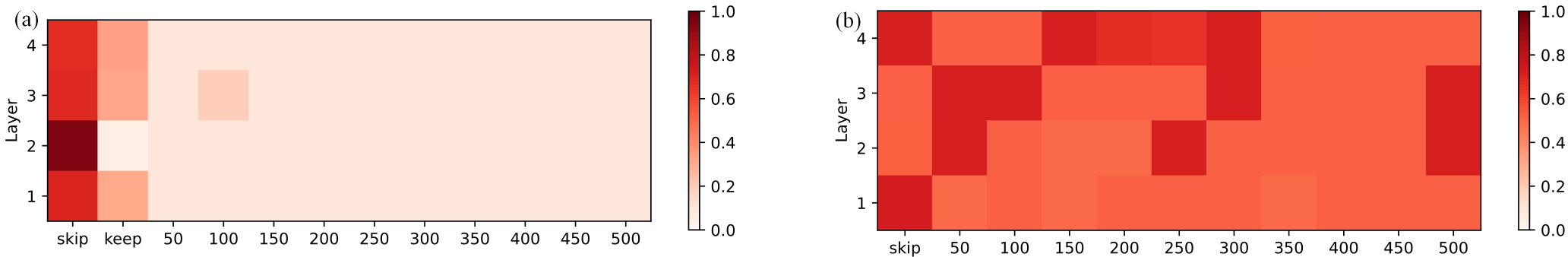}
	\caption{The visulization of architecture-related weights of (a) NAS-PINN and (b) NAS-PINNv2 in the case of arc plasma of Ar with RK-PINN framework}
	\label{fig:fig5}
\end{figure}

\begin{figure}
	\centering
	\includegraphics[width=17cm]{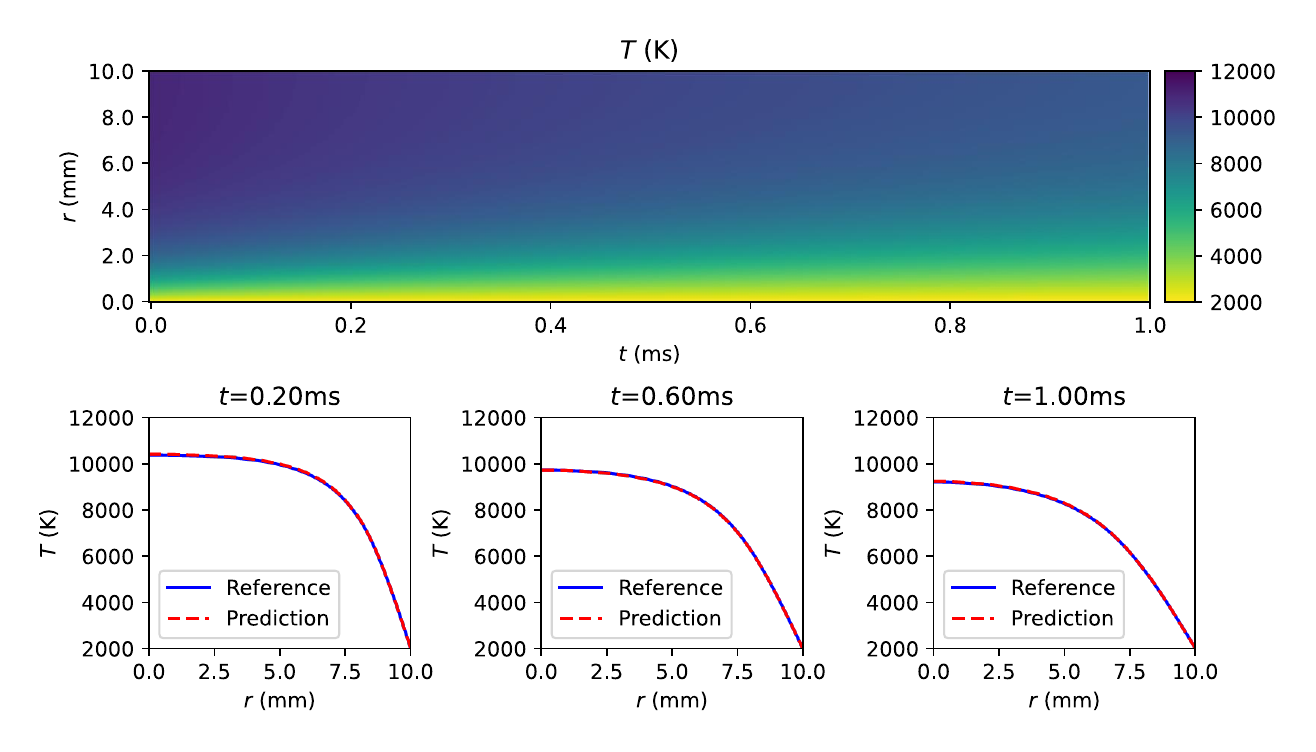}
	\caption{Prediction of the arc temperature in arc plasma of Ar by NAS-PINNv2 with CS-PINN framework}
	\label{fig:fig6}
\end{figure}

\begin{figure}
	\centering
	\includegraphics[width=17cm]{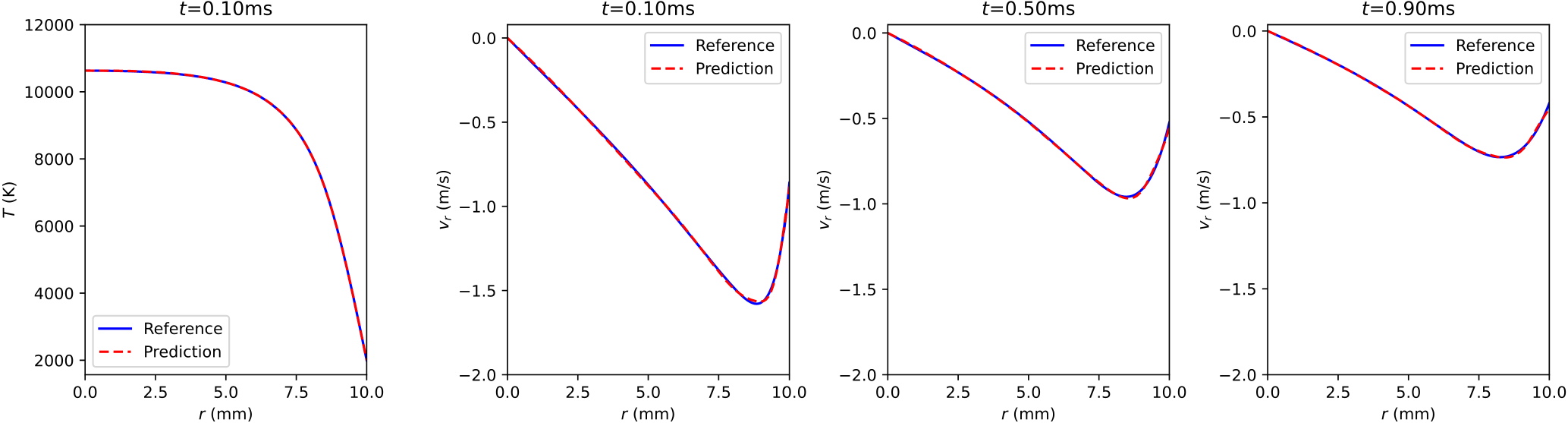}
	\caption{Prediction of the radial velocity in arc plasma of Ar by NAS-PINNv2 with RK-PINN framework}
	\label{fig:fig7}
\end{figure}

\paragraph{}
In this case, the vanilla NAS-PINN begins to exhibit widespread failure. Figure \ref{fig:fig5} visualizes the architecture-related weights obtained by NAS-PINN and NAS-PINNv2 in the case of Ar arc plasma with RK-PINN framework at $t=0.1$ ms. It is clearly shown that the weights assigned to skip operation are much larger than other operations, which hinders the vanilla NAS-PINN from extracting a reasonable discrete architecture. While NAS-PINNv2, applying the sigmoid function to calculate the architecture-related weights, avoids the suppression on other operations caused by skip operation. Figure \ref{fig:fig6}, Figure \ref{fig:fig7} and Table \ref{tab:tab2} show the results of the Elenbaas-Heller equation with radial velocity with CS-PINN and RK-PINN frameworks at different timesteps. As for the achieved relative error with CS-PINN, which are not included in Table \ref{tab:tab2}, the vanilla NAS-PINN failed, while NAS-PINNv2 reach a relative $L^2$ error as low as $2.14\times10^{-5}$ and $5.50\times10^{-4}$ in arc plasma of Ar and $2.95\times10^{-4}$ and $1.02\times10^{-3}$ in arc plasma of SF$_6$ for $T$ and $v_r$, respectively.

\begin{table}[htbp]
	\centering
	\caption{Elenbaas-Heller equation with radial velocity in arc plasma of Ar: relative $L^2$ error at different time steps with RK-PINN framework}
	\label{tab:tab2}
	\begin{tabular}{|l|l|l|l|l|}
\hline
$t$(ms)                  & Architecture name & Architecture                                & Relative $L^2$ error of $T$ & Relative $L^2$ error of $v_r$ \\ \hline
\multirow{3}{*}{$t=0.1$} & NAS-PINNv2        & {[}1, 0+150, 50+100, 50+250, 2{]}           & $4.15\times10^{-4}$              & $3.39\times10^{-3}$               \\ \cline{2-5} 
                       & NAS-PINN          & failed                                      & /                      & /                       \\ \cline{2-5} 
                       & Giant             & {[}1, 500$\times$4, 2{]}                           & $1.01\times10^{-3}$              & $6.61\times10^{-3}$               \\ \hline
\multirow{3}{*}{$t=0.5$} & NAS- PINNv2       & {[}1, 0+250, 50+400, 350+400, 150+350, 2{]} & $6.09\times10^{-4}$              & $8.51\times10^{-3}$               \\ \cline{2-5} 
                       & NAS-PINN          & failed                                      & /                      & /                       \\ \cline{2-5} 
                       & Giant             & {[}1, 500$\times$4, 2{]}                           & $6.48\times10^{-4}$              & $1.44\times10^{-2}$               \\ \hline
\multirow{3}{*}{$t=0.9$} & NAS-PINNv2        & {[}1, 400, 400, 250, 2{]}                   & $7.29\times10^{-4}$              & $8.78\times10^{-3}$               \\ \cline{2-5} 
                       & NAS-PINN          & failed                                      & /                      & /                       \\ \cline{2-5} 
                       & Giant             & {[}1, 500$\times$4, 2{]}                           & $1.62\times10^{-3}$              & $1.17\times10^{-2}$               \\ \hline
	\end{tabular}
\end{table}

\paragraph{}
NAS-PINNv2 can always discover a reasonable neural architecture facing the complex equations, while the vanilla NAS-PINN cannot discover any feasible neural architecture in this case. The neural architectures discovered by NAS-PINNv2 become more complicated as the scenario gets more complex. Multiple neuron numbers are required for a single hidden layer (expressed as A$+$B in Table \ref{tab:tab2}) and residual connections (expressed as 0 in Table \ref{tab:tab2}) begin to emerge in the final architecture. It is interesting to find that more than one neuron number can be selected for a single layer, which is rather rare in fully connected networks but more common in convolutional networks. This finding indicates that despite their relatively simple nature, fully connected networks can adopt more flexible design principles. Moreover, the NAS-PINNv2-searched architectures can achieve a lower relative $L^2$ error for both the arc temperature and the radial velocity. It can typically reduce the error by half for the radial velocity, which is empirically a challenging term to solve, compared to Giant. Table \ref{tab:tab2} also indicates a pattern that as the time step increases, NAS-PINNv2 tends to select a larger number of neurons per layer, which aligns with the intuition that equations with larger time steps are more challenging to solve.

\subsection{Drift-diffusion-Poisson equation for 1D DC corona discharge plasma}
\label{sec:sec4.3}
\paragraph{}
Corona discharge plasma is a common atmospheric pressure discharge phenomenon driven by a high voltage. To describe the transport characteristics of electrons and ions under the coupling of electric fields in a corona discharge plasma, the drift-diffusion-Poisson equation is introduced. Considering a 1D DC corona discharge plasma of argon where negative ions and the electron attachment process are neglected, the drift-diffusion-Poisson equation can be written as \cite{cite36}:

\begin{equation}
\label{equ:equ18}
	\begin{array}{l}
   \frac{\partial {{n}_{e}}}{\partial t}=\nabla \cdot \left( {{\mu }_{e}}\vec{E}{{n}_{e}}+{{D}_{e}}\cdot \nabla {{n}_{e}} \right)+\alpha {{n}_{e}}\left| {{\mu }_{e}}\vec{E} \right|, \\ 
  \frac{\partial {{n}_{p}}}{\partial t}=\nabla \cdot \left( -{{\mu }_{p}}\vec{E}{{n}_{p}}+{{D}_{p}}\cdot \nabla {{n}_{p}} \right)+\alpha {{n}_{e}}\left| {{\mu }_{e}}\vec{E} \right|, \\ 
  {{\nabla }^{2}}V=-\frac{e\left( {{n}_{p}}-{{n}_{e}} \right)}{\varepsilon } \\ 
\end{array}
\end{equation}

where $n_e$, $\mu_e$ and $D_e$ are the number density, mobility and diffusion coefficient of electrons, respectively, $n_p$, $\mu_p$ and $D_p$ are the corresponding quantities of positive ions, $E$ is the electric field strength, $\alpha$ is the ionization coefficient, $V$ is the applied DC voltage, $e$ is the elementary charge, $\varepsilon$ is the permittivity.

\paragraph{}
Similar to the simplification made in Zhong’s work \cite{cite21}, since electrons move much faster than heavy particles in a corona discharge, the ion density is considered time-independent, thus eliminating the need to solve np. The initial number densities of electrons and positive ions can be written as:

\begin{equation}
\label{equ:equ19}
	{{n}_{e(p)}}({{t}_{0}})={{n}_{\max }}\left( {{k}_{0}}+\exp \left( -{{r}^{2}}/2s_{0}^{2} \right) \right)
\end{equation}

where $n_{\max}$ is the maximum number density, $k_0$ and $s_0$ are the parameters to control the function shape. In this case, they are set to be $10^{15}$ m$^{-3}$, 0.001 and 0.25 cm, respectively.

\paragraph{}
As for the boundary conditions, a DC voltage of -10 kV is applied to the cathode, while the anode, positioned 1 cm apart, is grounded. Additionally, to describe the secondary electrons generated by the impact of positive ions on the cathode surface, the electron flux $\Gamma_e$ is introduced, which satisfies a zero-flux boundary condition at the anode. The electron flux can be calculated according to:

\begin{equation}
\label{equ:equ20}
	{{\Gamma }_{e}}=\gamma {{n}_{p}}{{\mu }_{p}}\left| {\vec{E}} \right|
\end{equation}

where $\gamma$ is the secondary electron emission coefficient and for a corona discharge in argon, it is set to 0.066.

\begin{figure}
	\centering
	\includegraphics[width=15cm]{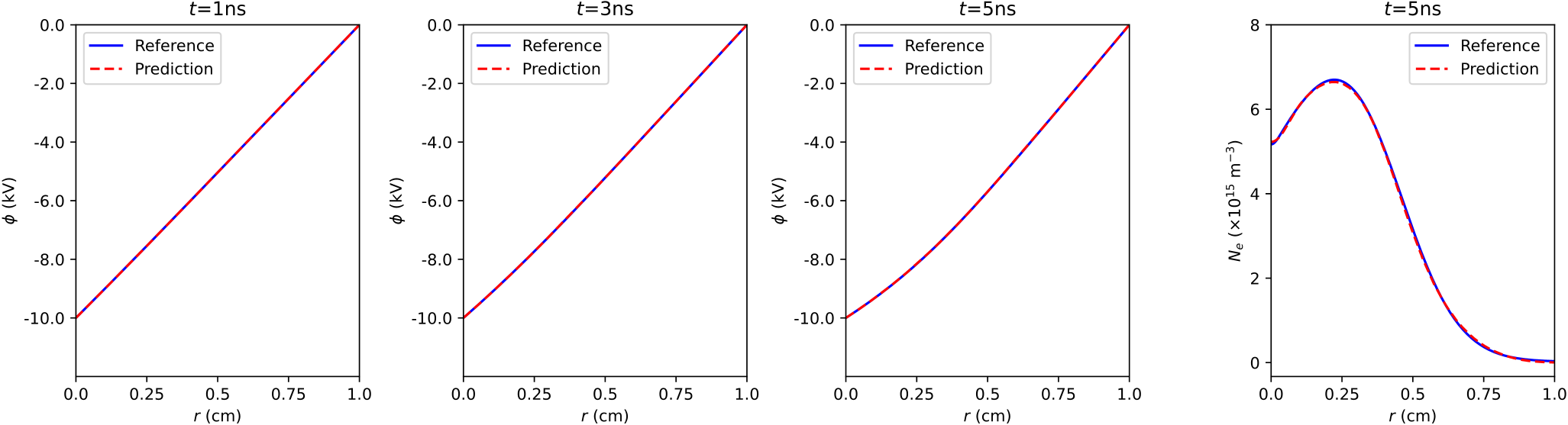}
	\caption{Prediction of $\varphi$ by NAS-PINNv2 with RK-PINN framework}
	\label{fig:fig8}
\end{figure}

\begin{figure}
	\centering
	\includegraphics[width=15cm]{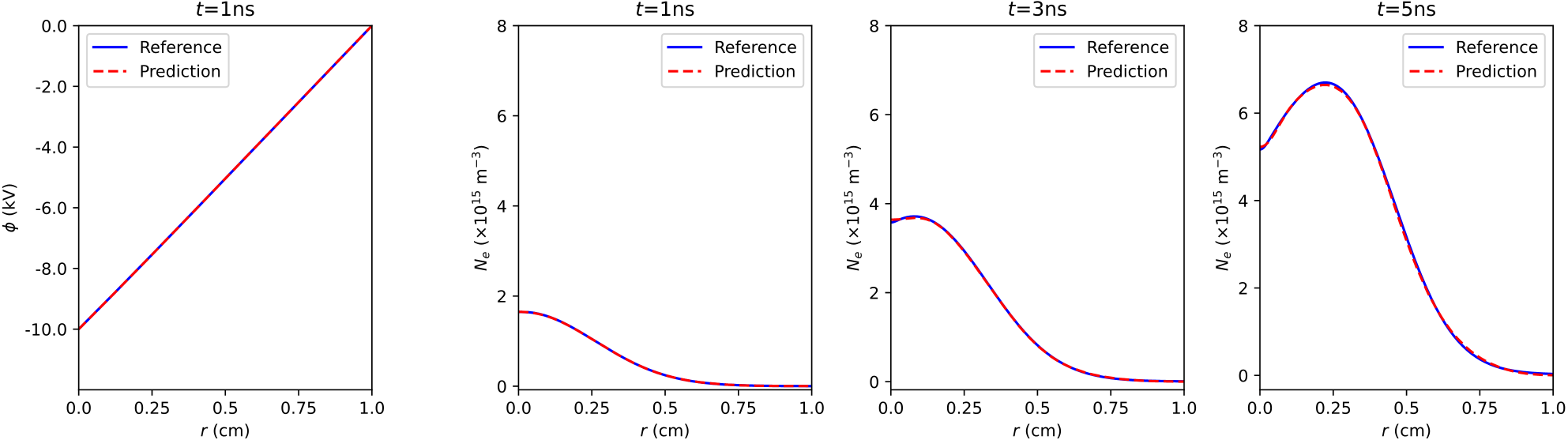}
	\caption{Prediction of $N_e$ by NAS-PINNv2 with RK-PINN framework}
	\label{fig:fig9}
\end{figure}

\paragraph{}
Since the corona discharge can evolve rapidly under high voltage, the numerical results mainly focus on the neural architecture of RK-PINN and the stage $q$ is set to 300. The investigated time span ranges from 0 to 6 ns and the reference solutions are obtained through an explicit time-advancing method based on Chebfun \cite{cite37}. The search space is set to be a neural network with up to 5 hidden layers, and the candidate neuron numbers for each layer range from 50 to 400 in increments of 50. For the inner loop, RK-PINN is trained with 500 collocation points uniformly sampled along the spatial coordinate. For the outer loop, 1000 collocation points are used to search for the optimal neural architecture, and the discovered neural architectures are trained from scratch. Table \ref{tab:tab3}, Figure \ref{fig:fig8} and Figure \ref{fig:fig9} give the results of predicted $\varphi$ and $N_e$ at different time steps.

\begin{table}[htbp]
	\centering
	\caption{Drift-diffusion-Poisson equation: relative $L^2$ error at different time steps}
	\label{tab:tab3}
	\begin{tabular}{|l|l|l|l|l|}
\hline
$t$(ns)                & Architecture name & Architecture                      & Relative $L^2$ error of $\varphi$ & Relative $L^2$ error of $N_e$ \\ \hline
\multirow{3}{*}{$t=1$} & NAS-PINNv2        & {[}1, 100, 250, 50, 100, 50, 2{]} & $3.45\times10^{-5}$              & $1.85\times10^{-3}$               \\ \cline{2-5} 
                     & NAS-PINN          & {[}1, 350, 400, 350, 2{]}         & $7.63\times10^{-5}$              & $2.83\times10^{-3}$               \\ \cline{2-5} 
                     & Giant             & {[}1, 400$\times$5, 2{]}                 & $7.13\times10^{-5}$              & $9.52\times10^{-3}$               \\ \hline
\multirow{3}{*}{$t=3$} & NAS- PINNv2       & {[}1, 50, 100, 50, 50, 50, 2{]}   & $2.97\times10^{-5}$              & $3.84\times10^{-3}$               \\ \cline{2-5} 
                     & NAS-PINN          & {[}1, 200, 400, 2{]}              & $4.46\times10^{-5}$              & $6.49\times10^{-3}$               \\ \cline{2-5} 
                     & Giant             & {[}1, 400$\times$5, 2{]}                 & $1.51\times10^{-4}$              & $4.39\times10^{-3}$               \\ \hline
\multirow{3}{*}{$t=5$} & NAS-PINNv2        & {[}1, 100, 100, 100, 50, 50, 2{]} & $1.27\times10^{-4}$              & $6.33\times10^{-3}$               \\ \cline{2-5} 
                     & NAS-PINN          & {[}1, 100, 200, 400, 400, 2{]}    & $1.52\times10^{-4}$              & $9.09\times10^{-3}$               \\ \cline{2-5} 
                     & Giant             & {[}1, 400$\times$5, 2{]}                 & $1.69\times10^{-4}$              & $7.927\times10^{-3}$               \\ \hline
	\end{tabular}
\end{table}

\paragraph{}
In this case, both NAS-PINNv2 and the vanilla NAS-PINN are capable of discovering reasonable architectures. However, NAS-PINN may achieve a higher relative error at certain time steps than the reference architecture Giant, while NAS-PINNv2 can keep a lowest relative error in all the cases. Moreover, the neural architectures searched by NAS-PINNv2 generally apply smaller numbers of neurons, verifying again that a larger neural network does not necessarily bring out a better performance.

\subsection{Boltzmann equation for electron transport in weakly ionized plasma}
\label{sec:sec4.4}
\paragraph{}
The Boltzmann equation provides a statistical representation of the distribution function of velocity or energy for species (such as electrons, ions or neutral particles) \cite{cite38}. In weakly ionized plasmas where electron transport dominates, the Boltzmann equation is widely used to solve for the electron energy distribution function (EEDF). Considering the Boltzmann equation in a polar coordinate, it can be expressed as \cite{cite39}:

\begin{equation}
\label{equ:equ21}
	\begin{array}{l}
   {{{\bar{R}}}_{i}}f\left( v,\theta  \right)+\frac{eE}{m}\left( \cos \theta \frac{\partial f\left( v,\theta  \right)}{\partial v}-\frac{\sin \theta }{v}\frac{\partial f\left( v,\theta  \right)}{\partial \theta } \right)-{{\left( \frac{\partial f\left( v,\theta  \right)}{\partial t} \right)}_{\text{coll}}}=0, \\ 
  {{{\bar{R}}}_{i}}=2\pi \int_{0}^{\infty }{\int_{0}^{\pi }{N\left( {{Q}_{i}}\left( v \right)-{{Q}_{a}}\left( v \right) \right)v}}f\left( v,\theta  \right){{v}^{2}}\sin \theta d\theta dv, \\ 
  {{\left( \frac{\partial f\left( v,\theta  \right)}{\partial t} \right)}_{\text{coll}}}=\frac{1}{2}N{{Q}_{el}}\left( {{v}_{el}} \right)v{{\left( 1+2m/M \right)}^{2}} \\ 
  \text{                         }\times \int_{0}^{\pi }{f\left( {{v}_{el}},\theta  \right)\sin \theta d\theta +\frac{1}{2}N{{Q}_{ex}}\left( {{v}_{ex}} \right)\frac{v_{ex}^{2}}{v}} \\ 
  \text{                         }\times \int_{0}^{\pi }{f\left( {{v}_{ex}},\theta  \right)\sin \theta d\theta +\frac{1}{2\Delta }N{{Q}_{i}}\left( {{v}_{i1}} \right)\frac{v_{i1}^{2}}{v}} \\ 
  \text{                         }\times \int_{0}^{\pi }{f\left( {{v}_{i1}},\theta  \right)\sin \theta d\theta +\frac{1}{2\left( 1-\Delta  \right)}N{{Q}_{i}}\left( {{v}_{i2}} \right)\frac{v_{i2}^{2}}{v}} \\ 
  \text{                         }\times \int_{0}^{\pi }{f\left( {{v}_{i2}},\theta  \right)\sin \theta d\theta -N{{Q}_{T}}\left( v \right)vf\left( v,\theta  \right)} \\ 
\end{array}
\end{equation}

where $\bar{R}_i$ is the effective ionization frequency, $v$ is the electron velocity, $\theta$ is the polar angle opposite to the electric field direction, $E$ is the applied electric field, $e$ is the elementary charge, $m$ and $M$ are the mass of electrons and molecules, respectively, $Q_i$, $Q_a$, $Q_{el}$ and $Q_{ex}$ are the electron-impact cross sections corresponding to the processes of ionization, electron attachment, elastic collision and excitation, respectively, $Q_T$ is the total electron-impact cross section, $v_{el}$, $v_{ex}$, $v_{i1}$ and $v_{i2}$ are the electron velocities associated with elastic, excitation and ionization collision processes, respectively, $\Delta$ is the ratio of energy sharing between two electrons during a ionization collision, which is typically set to 0.5.

\paragraph{}
In this case, CS-PINN is applied to solve the given Boltzmann equation to acquire $f(v, \theta)$ and the reference solutions are obtained by Bolsig$+$ \cite{cite39}. The maximum number of hidden layers in the search space is set to 4, and the candidate neuron numbers per layer range from 200 to 600 with an interval of 50. In the inner training phase, 200 points along the velocity and angle axis are uniformly sampled, while in the outer architecture search phase, 80 points are sampled. All the architectures are trained from scratch before tested.

\begin{figure}
	\centering
	\includegraphics[width=14cm]{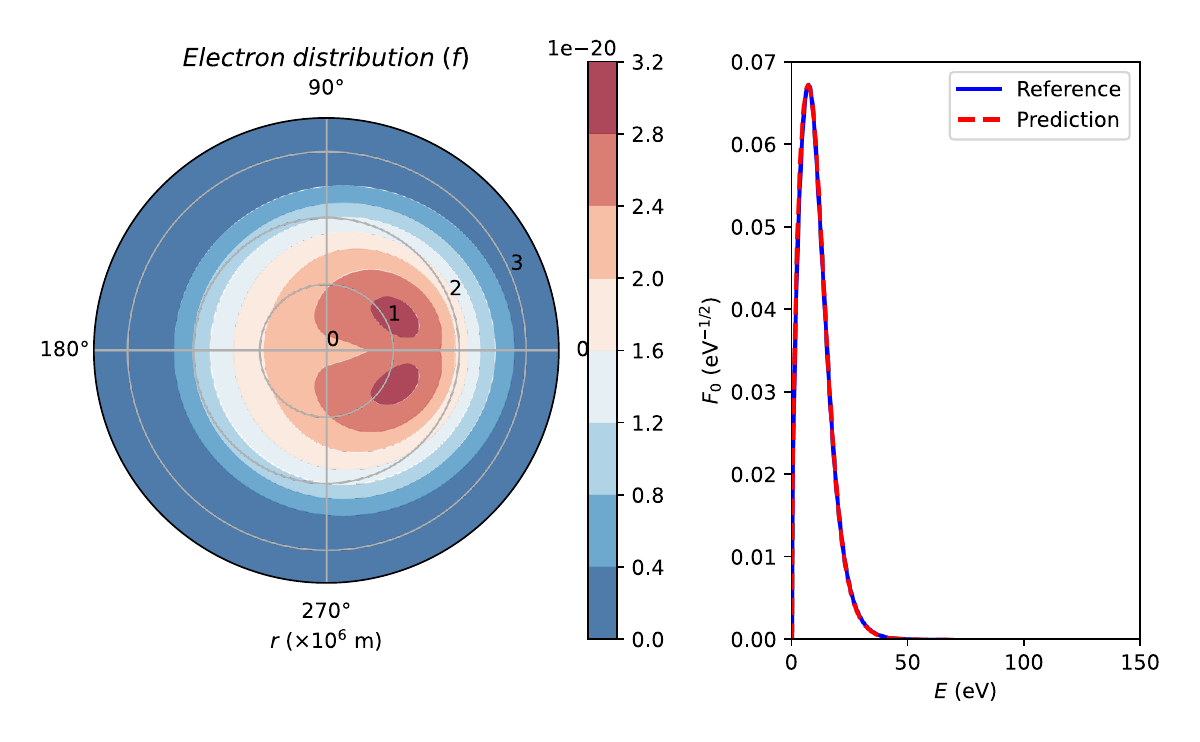}
	\caption{Prediction of electron distribution and EEDF by NAS-PINNv2}
	\label{fig:fig10}
\end{figure}

\paragraph{}
Figure \ref{fig:fig10} illustrates the electron distribution in an argon plasma predicted by NAS-PINNv2-searched architecture at a reduced electric field strength E/N of 500Td. Table \ref{tab:tab4} shows the relative $L^2$ error achieved by different neural architectures. Although the relative error achieved by different architectures do not vary much, the neural architecture discovered by NAS-PINNv2 obviously has fewer parameters than the other ones.

\begin{table}[htbp]
	\centering
	\caption{Boltzmann equation: relative $L^2$ error at different architectures}
	\label{tab:tab4}
	\begin{tabular}{|l|l|l|}
\hline
Architecture name & Architecture                   & Relative $L^2$ error of $F_0$ \\ \hline
NAS-PINNv2        & {[}2, 200, 450, 2{]}           & $2.24\times10^{-2}$               \\ \hline
NAS- PINN         & {[}2, 200, 600, 350, 600, 2{]} & $2.94\times10^{-2}$               \\ \hline
Giant             & {[}2, 600$\times$4, 2{]}              & $2.90\times10^{-2}$               \\ \hline
\end{tabular}
\end{table}

\section{CONCLUSIONS}
\label{sec:sec5}
\paragraph{}
In this paper, we propose an improved version of NAS-PINN to deal with the complex scenarios in LTP simulations. By analyzing the failure phenomenon of NAS-PINN in the process of solving plasma governing PDE, the exclusive competition of skip operation is investigated. Based on the idea of eliminating such competition, the sigmoid function is applied to calculate the architecture-related weights and a new loss term is introduced to constrain the architecture-related weights.

\paragraph{}
Through various numerical experiments including the Elenbaas-Heller equation, the drift-diffusion-Poisson equation and the Boltzmann equation, we verify the effectiveness of NAS-PINNv2 in complex LTP simulation scenarios with variable equation coefficients and strong nonlinearity. The results prove that NAS-PINNv2 can make a balance between skip operation and other operations, ensuring the proper search of neural architectures. Moreover, the occurrence of multiple neuron numbers for a single hidden layer reminds us that despite the relatively simple nature of fully connected network structures, we can still design more sophisticated architectures by modifying the connection of the data flows. The neural architecture search results can also help better understand and utilize residual connections.

\paragraph{}
Nevertheless, NAS-PINNv2 may yield a result not so satisfying due to the discretization from relaxed search space and the introduction of the sigmoid function exacerbates such discrepancy to some extent. Designing a new loss term to constrain the number of increasing architecture-related weights may potentially help mitigate this issue, and that can be further investigated in the future.

\section*{Acknowledgments}
\label{sec:acknowledgments}
\paragraph{}
This work was supported in part by the National Natural Science Foundation of China (92470102) and the Natural Science Foundation of Jiangsu Province (BK20231427).

\bibliographystyle{unsrt}


\end{document}